\documentclass[aps,twocolumn,twoside,secnumarabic,balancelastpage,amsmath,amssymb,hyperref=pdftex,superscriptaddress 
]{revtex4}

\usepackage{graphics}      
\usepackage[pdftex]{graphicx}      
\usepackage{longtable}     
\usepackage{bm}            
\usepackage{verbatim}
\usepackage{mathtools}
\usepackage{color}
\usepackage[colorlinks=true]{hyperref}
\begin{document}

\title{Hidden fluctuations close to a quantum bicritical point}
\date{\today}
	
\author{Corentin Morice}
\affiliation{Cavendish Laboratory, University of Cambridge, Cambridge CB3 0HE, United Kingdom}
\author{Premala Chandra}
\affiliation{Dept. of Physics and Astronomy, Rutgers University, Piscataway, New Jersey 08854, USA }
\author{Stephen E. Rowley}
\affiliation{Cavendish Laboratory, University of Cambridge, Cambridge CB3 0HE, United Kingdom}
\affiliation{Centro Brasileiro de Pesquisas Fisicas, Rio de Janeiro, 22290-180, Brasil}
\author{Gilbert Lonzarich}
\affiliation{Cavendish Laboratory, University of Cambridge, Cambridge CB3 0HE, United Kingdom}
\author{Siddharth S. Saxena}
\affiliation{Cavendish Laboratory, University of Cambridge, Cambridge CB3 0HE, United Kingdom}
\affiliation{National University of Science and Technology “MISiS”, Leninsky Prospekt 4, Moscow 119049, Russia}
\date{\today}

\begin{abstract}
Here we present an alternative approach for the description of quantum critical fluctuations. These are described by Langevin random fields, which are then related to the susceptibility using the fluctuation-dissipation theorem. We use this approach to characterise the physical properties arising in the vicinity of two coupled quantum phase transitions. We consider a phenomenological model based on two scalar order parameter fields locally coupled biquadratically and having a common quantum critical point as a function of a quantum tuning parameter such as pressure or magnetic field. A self-consistent treatment shows that the uniform static susceptibilities of the two order parameter fields have the same qualitative form at low temperature even where the forms are different in the absence of the biquadratic coupling.
\end{abstract}

\maketitle

Phase transitions in the low temperature limit can exhibit anomalous quantum critical behaviour and, at sufficiently low temperatures, novel intervening phases that in many cases have been surprising and challenging to describe theoretically (see e.g., \cite{Vojt2003, Monthoux2007, Sachdev2011, Wang2015}). One of the simplest examples involves a continuous phase transition between a displacive ferroelectric to a quantum paraelectric state as a function of pressure or via isotopic substitution at low temperatures \cite{Rechester1971, Khmelnitskii1973, Roussev2003, Das2009, Palova2011, Rowley2014, Rowley2016}. Here we wish to consider an alternative to current theories describing such transitions \cite{Lonzarich1985, Rowley2014} and its use in cases in which two order parameter fields are coupled (see, e.g., \cite{Calabrese2003, Millis2010, She2010, Das2012, Tokiwa2013, Oliver2015}). Our treatment will be specific to cases where each of the two order parameters is symmetry-breaking so that the lowest order terms coupling them is biquadratic; an important example is a multiferroic where the magnetisation and the polarisation break time-reversal and inversion symmetries respectively.

\begin{figure}[t]
\centering
\includegraphics[width=8.5cm]{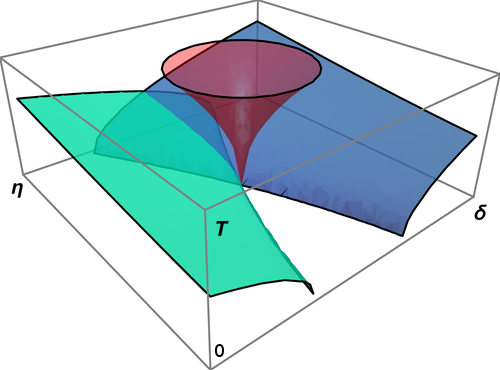}
\caption{Schematic phase diagram of the phase transition temperatures (blue and cyan) corresponding to two order parameters as a function of quantum tuning variables $\delta$ and $\eta$.  The quantum critical behaviour discussed in the text may be expected to be relevant within a narrow conical surface (illustrated in red) emanating from the bicritical point.  We assume that the transitions are second order but analyses similar to that given in the text have also been given for transitions that are weakly first order (see, e. g., \cite{She2010, Grigera2001}).}
\label{3DphaseDiagram}
\end{figure}

Our aim is to treat the problem of coupled symmetry-breaking fields in a simple and physically transparent way using a generalization of the Langevin random field concept.  We first discuss the possible range of validity of this approach for a single scalar order parameter field and then apply the approach to the case of two scalar fields to infer, in particular, how the quantum critical behaviour of the uniform static susceptibilities of the two fields might be expected to change in the presence of a local biquadratic coupling. For simplicity we assume that the phase transitions are second order and that they have a common quantum critical point, i.e., a quantum bicritical point, as a function of a tuning parameter such as pressure, dopant concentration, magnetic field, or electric field (Figure \ref{3DphaseDiagram}).  Importantly, we consider self-consistency only at the level of the static and uniform parameters of the field description. The possible limitations of this and other assumptions will be discussed.

\section{Biquadratically coupled order parameter fields}
\label{Biquadratically coupled order parameter fields}

We wish to consider two coupled order parameter scalar fields, $A$ and $B$, that in general fluctuate in space and time. The thermodynamic behaviour including the role of quantum dynamics is formally described in terms of an effective Lagrangian density in space and imaginary time. Here we adopt a simpler description in terms of a free energy density $\mathcal{F}$, analogous to that in the Ginzburg-Landau theory, and incorporate quantum dynamics by employing the fluctuation-dissipation theorem. Moreover, the effects of the self interaction and mutual interaction of the fields will be treated by means of a random field approach, analogous to that used to describe Brownian motion, which is mathematically efficient and physically transparent, yet in keeping with more formal approaches in the cases we shall consider here.

We assume that the coarse-graining of the fields is such that $\mathcal{F}$ can be expanded in a low-order power series in the amplitudes and spatial gradients of the fields with essentially temperature independent expansion parameters. In this case the temperature dependence of the uniform and static susceptibilities can be inferred from the self-consistent effects of the spontaneous fluctuations in the fields themselves. For simplicity, we begin with an expansion of $\mathcal{F}$ neglecting the gradient terms and assuming for definiteness a biquadratic coupling of the fields:
\begin{align*}
\mathcal{F} [A,B] = &\frac{a_1}{2}A^2+\frac{a_2}{4}A^4 + \frac{b_1}{2}B^2 + \frac{b_2}{4}B^4 + \frac{g}{2}A^2B^2 \\ &- A H_A - B H_B
\end{align*}
where $a_{1,2}$ and $b_{1,2}$ are expansion coefficients ($a_{2}>0$, $b_{2}>0$ as required for second order transitions), $g$ is the biquadratic coupling parameter, and $H_A$ and $H_B$ are the fields conjugate to the order parameters $A$ and $B$, respectively. We note that including terms second order in the gradients of the fields $\mathcal{F}$ would correspond to a Ginzburg-Landau free energy for the description of critical behaviour in the regime where $a_{1}$ and $b_{1}$ tend to vanish. The effects of the omitted gradient terms in $\mathcal{F}$ will be included below where needed.

Our first goal is to find the average values of the order parameter fields $\left\langle A \right\rangle$ and $\left\langle B \right\rangle$ that are stabilized by the conjugate fields $H_A$ and $H_B$, given the probability distribution for $A$ and $B$ defined through $\mathcal{F}$.  To do this we consider a simplified approach starting with the most probable values of order parameter fields obtained by minimizing $\mathcal{F}$ with respect to $A$ and $B$, which yields:
\begin{align*}
H_A = \left( a_1 + g B^2 \right) A + a_2 A^3\\
H_B = \left( b_1 + g A^2 \right) B + b_2 B^3\\
\end{align*}
We introduce the effects of quantum and thermal fluctuations by expressing $A$ and $B$ as sums of averages in the presence of the applied fields plus fluctuations about the average imagined to arise from Langevin random fields of zero mean. These random fields are the analogues of the random field introduced in Langevin's description of Brownian motion \cite{Reif}. A microscopic justification for our approach may be given in terms of an action principle as described in \cite{Feynman1965}.

\section{Susceptibility of a single order parameter field}

We first consider the case of a single order parameter field, e.g.,  $A$, and return to coupled fields in section \ref{Temperature dependence of the susceptibilities of two biquadratically coupled order parameter fields}. Setting $g=0$, and substituting in the above equation of state $H_A \to H_A+h$ and $A \to \left\langle A \right\rangle + a$,  where the fluctuation fields $h$ and $a$ have zero mean, then by averaging the resulting equation over $h$ and $a$ we obtain in lowest order in the fluctuations in $a$:
\begin{align*}
H_A = \left( a_1 + 3 a_2 \left\langle a^2 \right\rangle \right) \left\langle A \right\rangle + a_2 \left\langle A \right\rangle^3
\end{align*}
This is the average equation of state that we seek. It yields an initial inverse susceptibility of the form:
\begin{align*}
\chi_A^{-1} &= a_1 + 3 a_2 \left\langle a^2 \right\rangle
\end{align*}
where $\left\langle a^2 \right\rangle$ is the variance of the fluctuations in $a$ in the limit of vanishing value of $\left\langle A \right\rangle$.  The variance of the fluctuations affects the susceptibility through the anharmonic (quartic) term in the free energy density.

The possible range of validity of this Langevin random field approach will be discussed further through examples in the following section.

\section{Temperature dependence of the variance of the fluctuations}
\label{Temperature dependence of the variance of the fluctuations}

The variance $\left\langle a^2 \right\rangle$ can be obtained in principle from the full free energy density including the gradient terms omitted above. However, $\left\langle a^2 \right\rangle$ may be obtained more simply by making use of the generalized Nyquist theorem, or fluctuation-dissipation theorem, that relates the variance to the dynamical wave vector dependent susceptibility, $\chi_A(q,\omega)$, and hence to the dynamical properties of the order parameter field. In particular we consider dynamics as defined by wave vector and frequency dependent susceptibilities at small $q$ and $\omega$ of the forms:
\begin{align}
\label{prop-susceptibility}
\chi_p(q,\omega) = \frac{1}{\chi_p^{-1}+c q^2 - \frac{\omega^2}{\gamma q^n}}\\
\label{diss-susceptibility}
\chi_d(q,\omega) = \frac{1}{\chi_d^{-1}+c q^2 - \frac{i\omega}{\gamma q^n}}
\end{align}
where $\chi_p^{-1} = \chi_p^{-1}(0,0)$, $\chi_d^{-1} = \chi_d^{-1}(0,0)$, and $c$, $\gamma$ and $n$ are constants. Equations \eqref{prop-susceptibility} and \eqref{diss-susceptibility} describe the dynamics characteristic, respectively, of undamped (propagating or non-dissipative) and overdamped (dissipative) harmonic oscillator modes labeled by the wave vector $q$.

For example, the dynamics of critical transverse polar optical modes at a ferroelectric quantum critical point are traditionally represented by equation \eqref{prop-susceptibility} with $n=0$, while the dynamics of critical spin fluctuation modes at a ferromagnetic quantum critical point in a metal have traditionally been represented by equation \eqref{diss-susceptibility} with $n=1$ (neglecting a possible logarithmic correction to the term quadratic in $q$ in three dimensions).

\subsection{Fluctuation-dissipation theorem}

Given that we assumed that the coupling between the two order parameters does not lead to qualitative changes in the dynamics, the variance $\left\langle a^2 \right\rangle$, for example, defined in section \ref{Biquadratically coupled order parameter fields} can be determined from the generalised susceptibility $\chi_A (q,\omega)$ via the generalized Nyquist theorem, i.e. the fluctuation dissipation theorem, in the form:
\begin{equation}
\left\langle a^2 \right\rangle = \frac{2}{\pi} \sum_{q<q_c} \int_0^{\omega_c} \left( \frac{1}{2} + n(\omega) \right) \text{Im} \left( \chi_A (q,\omega) \right) d\omega
\label{FP-theorem}
\end{equation}
where $q_c$ and $\omega_c$ are cut-offs defining the coarse graining of our field description (the field $A$ is assumed to be averaged over distances and times on the order of $\frac{1}{q_c}$ and $\frac{1}{\omega_c}$ respectively) and $n(\omega)=\frac{1}{e^{\omega/T}-1}$ is the Bose function. For simplicity we have taken units in which $\hbar=k_B=1$. Note that for the case of propagating modes, $\text{Im} \left( \chi_A (q,\omega) \right)$ is obtained from equation \eqref{prop-susceptibility} via the replacement of $\omega^2$ in the denominator by $\omega^2 + i \epsilon \omega$, where $\epsilon \rightarrow 0^+$.

The zero-point contribution to the variance defined by the term $\frac{1}{2} \text{Im} \left( \chi_A (q,\omega) \right)$ in the integrand of equation \eqref{FP-theorem} is more weakly temperature dependent than the second term and is assumed to renormalise the parameters of the field model. We focus on the thermal contribution defined by the term $n(\omega) \text{Im} \left( \chi_A (q,\omega) \right)$ in the integrand, which we assume determines the temperature dependence of $\left\langle a^2 \right\rangle$ and hence of $\chi_A (q,\omega)$. We define this thermal component of $\left\langle a^2 \right\rangle$ as:
\begin{equation}
\left\langle a^2 \right\rangle_T = \frac{2}{\pi} \sum_{q<q_c} \int_0^{\omega_c} n(\omega) \text{Im} \left( \chi_A (q,\omega) \right) d\omega
\label{thermal-component}
\end{equation}
It is dominated by contributions from wave vectors such that $\omega (q) < T$, where $\omega (q)$ is the characteristic frequency (the dispersion relation) of fluctuations of a Fourier component, or mode, of the field $A$ of wave vector $q$. For propagating and dissipative modes, $\omega (q)$ represents respectively the oscillation frequency spectrum and the relaxation frequency spectrum. We define the thermal cut-off wave vector $q_T$ by $\omega(q_T) \sim T$. If $q_T < q_c$ and $\omega(q) < \omega_c$ when $q<q_T$, equation \eqref{thermal-component} can be approximated as:
\begin{equation*}
\left\langle a^2 \right\rangle_T \approx \frac{2}{\pi} \sum_{q<q_T} \int_0^{\omega_c} \frac{T}{\omega} \text{Im} \left( \chi_A (q,\omega) \right) d\omega
\end{equation*}
Using the Kramers-Kronig relation between the imaginary and real parts of the wave vector and frequency dependent susceptibility, we finally obtain:
\begin{equation}
\left\langle a^2 \right\rangle_T \approx T \sum_{q<q_T} \chi_A (q)
\label{approximation}
\end{equation}
where $\chi_A (q) = \lim_{\omega \to 0} \text{Re} \left( \chi_A (q,\omega) \right)$. Note that the thermal cut-off wave vector $q_T$ that plays a central role in this analysis is essentially a generalised inverse de Broglie thermal wavelength.

\subsection{Temperature dependence}

If $\omega(q) \propto q^z$ at low $q$, where $z$ is the dynamical exponent, then $q_T \propto T^{1/z}$. For the model represented by equation \eqref{prop-susceptibility} we see that, in the most familiar case where $n=0$, at the quantum critical point we have $z=1$. On the other hand, for the model represented by equation \ref{diss-susceptibility} we see that at the quantum critical point $z=2+n$.

Moreover, in both the propagating (Equation \eqref{prop-susceptibility}) and dissipative (Equation \eqref{diss-susceptibility}) cases, the momentum-dependent susceptibility is:
\begin{equation}
\chi_A (q) = \frac{1}{c_A\left(\kappa^2 + q^2 \right) }
\label{A-susceptibility}
\end{equation}
where $\kappa = 1/\sqrt{c_A \chi_A}$ is the correlation wave vector, or inverse correlation length for the field $A$.

Using equation \eqref{approximation} together with equation \eqref{A-susceptibility} we find:
\begin{align*}
\left\langle a^2 \right\rangle_T \propto & T \sum_{q<q_T} \frac{1}{\kappa^2 + q^2 } \propto T \int_{0}^{q_T} \frac{q_T^{d-1}}{\kappa^2 + q^2 } dq
\\
\approx & T \int_{\kappa}^{q_T} \frac{q_T^{d-1}}{ q^2 } dq = \frac{T}{d-2} \left( q_T^{d-2} - \kappa^{d-2} \right)
\\
\propto & T^{\frac{d+z-2}{z}} \left( 1 - \left( \frac{\kappa}{q_T} \right)^{d-2} \right)
\end{align*}
At the quantum critical point, $\left\langle a^2 \right\rangle_T \propto \chi_A^{-1} \propto \kappa^2$, hence the previous equation leads to:
$$\left( \frac{\kappa}{q_T} \right)^{2} \propto T^{\frac{d+z-4}{z}} \left( 1 - \left( \frac{\kappa}{q_T} \right)^{d-2} \right)$$
We see that if $d+z > 4$ and $d>2$, $\lim_{T \to 0} \frac{\kappa}{q_T} = 0$, in which case in the low temperature limit:
$$\left\langle a^2 \right\rangle_T \propto \chi_A^{-1} \propto T^{\frac{d+z-2}{z}}$$
Thus we conclude that when the effective dimension $d+z$ exceeds the upper critical dimensions of 4 for our model, then the asymptotic quantum critical exponent is $e_A=\frac{d+z-2}{z}$. The dynamical properties of the field enter through the dynamical exponent $z$. These results can be confirmed by carrying out the full integrals in equation \eqref{FP-theorem} with the wave vector and frequency dependent susceptibilities defined by equations \eqref{prop-susceptibility} and \eqref{diss-susceptibility}.

For the case of dissipative modes, our approximation for the thermal variance holds only for $z>1$. For $z=1$ (e.g., for the non-critical case with $n=1$) the full solution of equation \eqref{FP-theorem} yields a quadratic temperature dependence of the thermal variance.

\subsection{Finite critical temperature}

It is interesting to compare the above with the standard analysis for the behaviour near to a finite critical temperature $T_c$. In that case we have from equations \eqref{approximation} and \eqref{A-susceptibility} and expanding in leading order in $\kappa^2$:
\begin{align}
\kappa^2 &= \frac{1}{c_A \chi_A} = \frac{a_1}{c_A} + \frac{3 a_2 T}{c_A^2} \sum_{q<q_T} \frac{1}{\kappa^2+q^2}
\\
&= \left( \frac{a_1}{c_A} + \frac{3 a_2 T}{c_A^2} \sum_{q<q_T} \frac{1}{q^2} \right) - \left( \frac{3 a_2 T}{c_A^2} \sum_{q<q_T} \frac{1}{q^4} \right) \kappa^2
\label{classical-expansion}
\\
&= \kappa_0^2 / \left( 1 + \frac{3 a_2 T}{c_A^2} \sum_{q<q_T} \frac{1}{q^4} \right)
\end{align}
where $\kappa_0^2$ is the first term on the right hand side of equation \eqref{classical-expansion}, which vanishes at $T_c$. Expanding $\kappa_0^2$, which is a smooth or analytic function of $T$ around a finite $T_c$, to leading order in $\Delta T=T-T_c$, we obtain the classical critical exponent of unity if $d>4$ so that the sum over $q$ in the denominator is finite. If $d<4$ we avoid the expansion in $\kappa^2$ and replace $-\frac{\kappa^2}{q^4}$ by the exact difference $\frac{1}{\kappa^2+q^2} - \frac{1}{q^2}$, and note that the classical exponent is expected to become invalid for $\Delta T$ below the Ginzburg scale $\Delta T_G$ roughly defined by the condition that the second term in equation \eqref{classical-expansion}, as corrected by using the exact difference, exceeds the first term.

\subsection{Examples for propagating and dissipative dynamics}

As an example involving propagating dynamics, consider a displacive ferroelectric quantum critical point. In the conventional description, the critical modes are transverse polar optical phonons whose energy gap vanishes at the quantum critical point. In this case from equation \eqref{prop-susceptibility} we have $n=0$, $z = 1$ and hence $e_A \to 2$ for $d=3$. However, note that since $d+z$ is just equal to the upper critical dimension, the condition $d+z>4$ given in the last section is not strictly fulfilled. This leads only to a weak logarithmic correction to the $T^2$ dependence of the inverse susceptibility, which may be difficult to observe in practice. Of greater importance are the effects of the long-range dipole-dipole interactions and the coupling of the critical polar modes to acoustic phonons that can lead to a breakdown of the $T^2$ temperature dependence at sufficiently low temperatures \cite{Rowley2014, Rowley2016}.

As an example involving dissipative dynamics, consider a magnetic quantum critical point for an itinerant electron ferromagnet. In the conventional description, the critical modes are dissipative spin fluctuations or paramagnons characterized by a relaxation spectrum the linear in $q$ component of which vanishes at the quantum critical point. In this case from equation \eqref{diss-susceptibility} we have $n=1$, $z \to 3$ and hence $e_A \to \frac{4}{3}$ for $d=3$. Here the condition $d+z>4$ is well fulfilled. However, gapless transverse spin fluctuations can lead to non-analytic corrections to the quadratic term in $q$ in the generalized susceptibility as well as to the quartic term in the order parameter in the free energy. These effects can change the nature of the quantum phase transition in an isotropic itinerant-electron ferrromagnet at sufficiently low temperatures and low magnetic fields.

We note that non-dissipative dynamics may be relevant not only to ferroelectrics but also to certain types of magnetic quantum phase transitions in insulators (see, e.g., \cite{Das2012, Sachdev2011}), while dissipative dynamics can arise not only in metals but also in insulators such as ferroelectric relaxors (see, e.g., \cite{Al-Zein2010}).

The examples given here along with numerous others show that the behaviour near to a quantum critical point can be subtle and complex even when the condition $d+z>4$ is met and only one critical field appears to be relevant at first sight. Despite these complexities, the predictions of the above simple models have been observed experimentally at least over limited ranges in temperatures and applied fields in examples of both displacive ferroelectrics and itinerant electron ferromagnets (see, e.g., \cite{Rowley2014} and references therein).

\section{Temperature dependence of the susceptibilities of two biquadratically coupled order parameter fields}
\label{Temperature dependence of the susceptibilities of two biquadratically coupled order parameter fields}

For the single order parameter field, the Langevin random field approach yields results in agreement with that obtained both by more formal techniques and by experiment for the case and conditions discussed in the last section.

Here we consider the predictions of the same approach for the case of two coupled fields as defined in section \ref{Biquadratically coupled order parameter fields}. We stress that as formulated here the approach considers self-consistency solely at the level of the static and uniform parameters of the field description.  Thus the results apply only if the dynamical properties of the fields are not modified qualitatively by the local biquadratic coupling introduced. Our experience with the simpler problems discussed in the previous section suggests that the simple analysis presented below may be relevant to some special cases and conditions (e.g., temperature ranges) of practical interest.

As in section \ref{Biquadratically coupled order parameter fields} we introduce the effects of spontaneous fluctuations by expressing $A$ and $B$ as sums of averages in the presence of the applied fields plus fluctuations about the average imagined to arise from Langevin random fields of zero mean.  Thus we take $A \to \left\langle A \right\rangle + a$ and $B \to \left\langle B \right\rangle + b$, respectively.  Averaging, as before, the equation of state in section \ref{Biquadratically coupled order parameter fields} in the limit where $\left\langle A \right\rangle$ and $\left\langle B \right\rangle$ are vanishingly small, keeping in mind that $\left\langle a \right\rangle = \left\langle b \right\rangle = 0$ and neglecting fluctuation terms higher than second order, we find for the inverse susceptibility for the order parameter $A$, for example:
\begin{align}
\chi_A^{-1} &= a_1 + g \left\langle b^2 \right\rangle + 3 a_2 \left\langle a^2 \right\rangle
\label{susceptibilityab}
\end{align}
The temperature dependence of the susceptibility $\chi_A$ is therefore given by the sum of the temperature dependence of the two last terms, weighted by the appropriate coefficients. The variances of $a$ and $b$ have zero-point and thermal contributions.  We assume, as in section \ref{Temperature dependence of the variance of the fluctuations}, that the former lead to renormalisations of the (weakly temperature dependent) parameters of the model and consider only the thermal contributions governed by the Bose function.  At low $T$ the two last terms in equation \eqref{susceptibilityab} lead to temperature dependences that can be expressed as:
$$ \chi_A^{-1} = c_0 + c_{AB} T^{e_B} + c_{AA} T^{e_A}$$
where the asymptotic quantum critical exponents $e_A$ and $e_B$ are governed by the dynamical properties of the fluctuations of the $A$ and $B$ fields, respectively, and $c_0$, $c_{AA}$ and $c_{AB}$ are constants; note that $c_0$ vanishes at the quantum critical point. The two temperature-dependent terms are equal at a crossover temperature:
$$ T_{cross_A} = \left( \frac{c_{AA}}{c_{AB}} \right)^{\frac{1}{e_B-e_A}}$$
For $T$ below $T_{cross_A}$ the  temperature dependence of the inverse susceptibility for the $A$ order parameter at the quantum critical point is given by:
$$\chi_A^{-1} \sim T^{min(e_A,e_B)}$$

By symmetry, this implies that at the quantum  bicritical point the asymptotic exponent is the same for $\chi_A$ and $\chi_B$ and is governed by the dynamics of the field that leads to the lower exponent. In this sense the dynamics of the other field with the higher exponent is hidden.  The difference in the temperature dependences of the two susceptibilities is the temperature at which the influence of the term with a higher exponent will become important. We note that where comparisons can be made our conclusions are consistent with those obtained previously (see, e.g., \cite{Das2012}) using a somewhat different approach.

\section{Conclusion}

The problem of two coupled fields considered in the section above is likely to be much more complex than that of the single order parameters considered in section \ref{Temperature dependence of the variance of the fluctuations}, which as we have discussed involve a number of subtleties themselves. However it is possible that, as for the single order parameter, the predictions for the coupled order parameters, namely that the susceptibilities for both order parameters may have the same critical exponent over a finite temperature range, may be relevant in certain materials under appropriate conditions. We note that quantum bicriticality induced by disorder \cite{Dalidovich1999, Demishev2013} and/or frustration \cite{Tokiwa2013} has been explored previously, but here we are exploring a different sort of quantum bicriticality due to the coupling of symmetry-breaking order parameters in a much simpler translationally-invariant system.

In the simple models we have considered here the relevant parameters are the dimensionality of space, $d$, and the dynamical exponent, $z$. Both of these parameters can differ for the two coupled order parameter fields leading to a number of distinct possibilities for quantum critical behaviour. The effective value of $d$ can differ for the two fields due to different strengths of interplanar interactions or due to dimensional cross-over as a function of temperature \cite{Lee1994,Valla2002}. The values of $z$ for the two coupled fields can also vary due to the extent of dissipation or, for example, in the case of magnetism, to the role of quantum precession \cite{Sachdev2011}. We also note that displacive ferroelectricity is not restricted to insulators, indeed it can arise in local form in metals as long as the Thomas-Fermi screening length is smaller than the ferroelectric correlation length. This can happen in carrier doped materials up to quite high carrier densities \cite{Hwang2010, Kolodiazhnyi2010}, and hence could potentially coexist with metallic magnetism. In such cases both weakly-dissipative dynamics in one order parameter field and strongly-dissipative dynamics in the other order parameter field, with generally quite different values of $z$, would seem to be relevant. The analysis presented here may provide some initial direction on how this diversity of behaviours of coupled order parameters might be at least partly probed by means of measurements of either one or the other of the two order parameter susceptibilities, whichever one happens to be the most convenient to observe.

The approach presented here, grounded on the generalization of familiar principles (Langevin random fields, Nyquist theorem and de Broglie thermal wavelength), may be helpful in developing intuitive phenomenological descriptions of these interesting and complex multicritical systems, as has already been demonstrated in elementary examples involving an order parameter field coupled to an auxiliary field \cite{Lonzarich1985, Rowley2014}.

The study of quantum multicritical points is still in its infancy but is expanding rapidly in materials that are not only of theoretical but also potentially of technological interest. These include materials involving both dielectric and magnetic instabilities, of the general kinds that arise, for example, in the hexagonal ferrites and perovskite oxides structure types, such as EuTiO$_3$ \cite{Rowley2016, Schiemer2016, Parisiades2015}. One of the most dramatic examples of the coexistence of multiple order parameter fields \cite{Badoux2015} of potentially comparable importance is found among the high temperature superconductors that continue to challenge conventional thinking.

\section{Acknowledgements}

We would like to thank Turan Birol, Richard Brierley, Christopher Hooley, Dmitry Kovrizhin, Peter Littlewood, Bruno Loureiro and Jasper van Wezel for fruitful discussions. We acknowledge support from EPSRC, Corpus Christi College, the National Science Foundation grant NSF DMR-1334428, the Increase Competitiveness Program of the Ministry of Education of the Russian Federation grant NUST MISiS K2-2017-024 and a CONFAP Newton grant. We are grateful to Trinity College Cambridge, the Cavendish Laboratory and the Aspen Center for Physics, supported by NSF PHY-1066293, for hospitality.

\bibliographystyle{apsrev4-1}
\bibliography{QuantumCriticality}

\end{document}